\documentclass[twocolumn,superscriptaddress,nofootinbib,floatfix,aps,prb,citeautoscript,longbibliography]{revtex4-2}
\usepackage[utf8]{inputenc}
\usepackage[T1]{fontenc}
\usepackage{lineno}
\usepackage{soul}
\usepackage[colorlinks=true,linkcolor=blue,citecolor=blue,urlcolor=blue]{hyperref}
\usepackage{graphicx}
\usepackage{xcolor}
\usepackage{array}
\usepackage{dcolumn}
\usepackage{bm}
\usepackage{multirow}
\usepackage[version=4]{mhchem}
\usepackage{cleveref}
\usepackage{orcidlink}

\begin{document}

\newcommand{\bochum}{Research Center Future Energy Materials and Systems of the University Alliance Ruhr and Interdisciplinary Centre for Advanced Materials Simulation, Ruhr University Bochum, Universit\"atsstraße 150, D-44801 Bochum, Germany}
\newcommand{\coimbra}{CFisUC, Department of Physics, University of Coimbra, Rua Larga, 3004-516 Coimbra, Portugal}
\newcommand{\mpihalle}{Max-Planck-Institut f\"ur Mikrostrukturphysik, Weinberg 2, D-06120 Halle, Germany}
\newcommand{\MLU}{Institut f\"ur Physik, Martin-Luther-Universit\"at Halle-Wittenberg, D-06099 Halle, Germany}
\newcommand{\ehu}{Fisika Aplikatua Saila, University of the Basque Country (UPV/EHU), Europa Plaza 1, 20018 Donostia/San Sebasti{\'a}n, Spain}
\newcommand{\cfm}{Centro de F{\'i}sica de Materiales (CFM-MPC), CSIC-UPV/EHU, Manuel de Lardizabal Pasealekua 5, 20018 Donostia/San Sebasti{\'a}n, Spain}
\newcommand{\dipc}{Donostia International Physics Center (DIPC), Manuel de Lardizabal Pasealekua 4, 20018 Donostia/San Sebasti{\'a}n, Spain}

\author{Kun Gao\orcidlink{0009-0007-6032-9683}}
\affiliation{\bochum} 
\author{Tiago F. T. Cerqueira\orcidlink{0000-0002-4147-8129}}
\affiliation{\coimbra}
\author{Antonio Sanna  \orcidlink{0000-0001-6114-9552}}
\affiliation{\mpihalle}
\affiliation{\MLU}
\author{Yue-Wen Fang\orcidlink{0000-0003-3674-7352}}
\affiliation{\cfm}
\author{{\DJ}or{\dj}e Dangi{\'c}\orcidlink{0000-0002-4639-184X}}
\affiliation{\cfm}
\affiliation{\ehu}
\author{Ion Errea\orcidlink{0000-0002-5719-6580}}
\affiliation{\cfm}
\affiliation{\ehu}
\affiliation{\dipc}
\author{Hai-Chen Wang\orcidlink{0000-0002-2892-5879}}
\affiliation{\bochum} 
\author{Silvana Botti\orcidlink{0000-0002-4920-2370}}
\affiliation{\bochum} 
\author{Miguel A. L. Marques\orcidlink{0000-0003-0170-8222}} 
\email{miguel.marques@rub.de}
\affiliation{\bochum} 

\date{\today}

\newcommand{\Tc}{\ensuremath{T_\textrm{c}}}
\newcommand{\Tcmac}{\ensuremath{T_\textrm{c}^\text{McMillan}}}
\newcommand{\Tcad}{\ensuremath{T_\textrm{c}^\text{Allen-Dynes}}}
\newcommand{\Tce}{\ensuremath{T_\textrm{c}^\text{Eliashberg}}}
\newcommand{\olog}{\ensuremath{\omega_\text{log}}}
\newcommand{\omax}{\ensuremath{\omega_\text{max}}}
\newcommand{\dosef}{\ensuremath{\text{DOS}({\text{E}_\text{F}}})}
\newcommand{\afo}{\ensuremath{\alpha^2F(\omega)}}
\newcommand{\la}{\ensuremath{\lambda}}

\title{The Maximum $T_\mathrm{c}$ of Conventional Superconductors at Ambient Pressure}

\begin{abstract}
The theoretical maximum critical temperature (\Tc{}) for conventional superconductors at ambient pressure remains a fundamental question in condensed matter physics. Through analysis of electron-phonon calculations for over 20\,000 metals, we critically examine this question. We find that while hydride metals can exhibit maximum phonon frequencies of more than 5000~K, the crucial logarithmic average frequency \olog\ rarely exceeds 1800~K. Our data reveals an inherent trade-off between \olog\ and the electron-phonon coupling constant $\lambda$, suggesting that the optimal Eliashberg function that maximizes \Tc\ is unphysical. Based on our calculations, we identify \ce{Li2AgH6} and its sibling \ce{Li2AuH6} as theoretical materials that likely approach the practical limit for conventional superconductivity at ambient pressure. Analysis of thermodynamic stability indicates that compounds with higher predicted \Tc\ values are increasingly unstable, making their synthesis challenging. While fundamental physical laws do not strictly limit \Tc\ to low-temperatures, our analysis suggests that achieving room-temperature conventional superconductivity at ambient pressure is extremely unlikely.
\end{abstract}

\maketitle

Since the groundbreaking discovery of superconductivity in mercury at 4.2~K in 1911~\cite{onnes1911superconductivity}, superconductors have revolutionized both fundamental physics and technological applications. These materials have enabled transformative technologies including magnetic resonance imaging machines, magnetic levitation trains, and ultra-sensitive quantum devices such as SQUIDs. The quest for high-temperature superconductivity remains one of the grand challenges of solid-state physics, as raising the critical temperature would unlock unprecedented possibilities in power transmission, quantum computing, and particle acceleration.

Superconducting materials are traditionally divided into conventional and unconventional. In the former, the topic of the current work, the mechanism responsible for superconductivity is the electron-phonon interaction, electrons form Cooper pairs in a singlet state, and the energy gap has s-wave symmetry. These compounds are well understood within the Bardeen, Cooper, and Schrieffer (BCS) theory~\cite{bardeen1957microscopic}, or its strong coupling generalization, Eliashberg theory~\cite{Eliashberg_1960}. 

In the 1960s and 1970s, it was believed that the maximum $\Tc$ for conventional superconductors would likely be in the range of 30--40~K. This was based on the observed properties of known superconducting materials and the theoretical limits imposed by the electron-phonon interaction. The argument was that to achieve a higher \Tc\ would require an unrealistically strong electron-phonon coupling or an exceptionally high density of states, both of which seemed improbable given the materials known at the time. These beliefs were supported by experimental data. For example, elemental superconductors like Pb and Nb have $\Tc=7.2$~K and $\Tc=9.25$~K, respectively~\cite{buzea2004assembling}, while more complex compounds like \ce{Nb3Sn} or \ce{Nb3Ge} reach $\Tc=17.9$~K and $\Tc=21.8$~K respectively~\cite{stewart2015superconductivity}.

Currently, the compound with record conventional superconductivity at room pressure is \ce{MgB2}, with a transition temperature of $\Tc=39$~K~\cite{nagamatsu2001superconductivity}. This compound exhibits two nearly noninteracting bands of different dimensionality~\cite{canfield2003magnesium}, leading to the coexistence of two energy gaps in the same material~\cite{liu2001beyond}. The idea of multiple gaps coexisting in a single superconductor had been explored previously~\cite{suhl1959bardeen,binnig1980two}, but \ce{MgB2} stands out as the first system where this phenomenon is so prominently and distinctly manifested~\cite{choi2002origin}.

An alternative approach to achieving high-temperature superconductivity involves the chemical precompression of the hydrogen sub-lattice in hydrides materials~\cite{flores2020perspective}. This led to groundbreaking discoveries of very high-\Tc\ superconductors, such as \ce{H3S}~\cite{li2014metallization,duan2014pressure,drozdov2015conventional}, \ce{LaH10}~\cite{liu2017potential,peng2017hydrogen,drozdov2019superconductivity,somayazulu2019evidence}, \ce{YH$_x$}~\cite{li2015pressure,kong2021superconductivity,troyan2019anomalous}, or even in ternary compounds such as \ce{LaBeH8}, \ce{(La,Y)H10}, \ce{(La,Ce)H9}, etc.~\cite{song2022potential,semenok2021superconductivity,song2023stoichiometric,bi2022giant,chen2023enhancement,semenok2022effect}. Unfortunately, all these systems require very high pressures, seriously limiting their application to technology.

A key question in this field is what is the upper limit to the critical temperature of superconductors~\cite{gor2018colloquium,Boeri_2019,Roadmap_RTS}. Based on the observation that the coupling constant depends mainly on the phonon frequencies, McMillan
 in 1968~\cite{mcmillan1968transition} already derived an approximate expression of $\Tc^\text{max}$. His result, however, only holds for a \textit{given class of materials} and therefore does not provide an absolute value for this quantity. Further insight can be obtained by maximizing \Tc\ given by McMillan's formula
\begin{equation}
  \Tc^\text{McMillan} = \frac{\olog}{1.20} \exp\left(-1.04\frac{1 + \lambda}{\lambda - \mu^* (1 + 0.62\lambda)}\right)
  \,,
\end{equation}
where $\mu^*$ is the Coulomb pseudopotential, $\lambda$ is the electron-phonon coupling constant
\begin{equation}
\lambda = 2 \int_0^{+ \infty} \frac{\alpha^2F(\omega)}{\omega} d\omega    
\end{equation}
and $\alpha^2F(\omega)$ is the Eliashberg spectral function, calculated from the electron-phonon coupling matrix elements.
With \olog\ we indicate the logarithmic average of the phonon frequencies:
\begin{equation}
\omega_{\text{log}} =\text{exp}\left[\frac{\lambda}{2}\int_0^{+ \infty}\ln(\omega)\frac{\alpha^2F(\omega)}{\omega}d\omega\right] \,.
  \end{equation}

By setting, for simplicity, the Coulomb pseudopotential $\mu^*$ to zero, one obtains that the maximum is attained at $\lambda\approx2$. However, McMillan's formula is only valid for values of $\lambda\lesssim1.5$, while no maximum value for \Tc\ exists in the original Eliashberg theory~\cite{gor2018colloquium}. It is also frequently argued that the value of $\lambda$ is limited, as the lattice becomes eventually unstable for very large values of the coupling constant. However, there are some experimentally known superconductors with very high values of $\lambda$, well above 2.0, both at ambient (such as Pb--Bi compounds~\cite{chen1971electron}) and under pressure~\cite{tanaka2017electron}.

Recent works estimated the maximum \Tc\ of conventional superconductivity from fundamental limits~\cite{ARXIV.2407.12922,Sadovskii2024,trachenko2024upper}. They all agree on a value of 300--600~K at ambient pressure, suggesting that superconductors may exist at ambient temperature. In the following, we critically look at the basic assumptions used in these estimates. This is done by analyzing experimentally-known superconductors and our calculations of the electron-phonon interaction and superconducting properties for more than 20\,000 metals~\cite{cerqueira2024searching,cerqueira2024sampling}. This is by far the largest dataset available to date with calculated superconducting properties, and it includes metals with a large variety of chemical elements and crystal structures. Although it is not comprehensive, it is rather extensive, giving us a very good overview of conventional superconductivity in materials space.

\begin{figure}[tb]
    \centering
    \includegraphics[width=0.9\columnwidth]{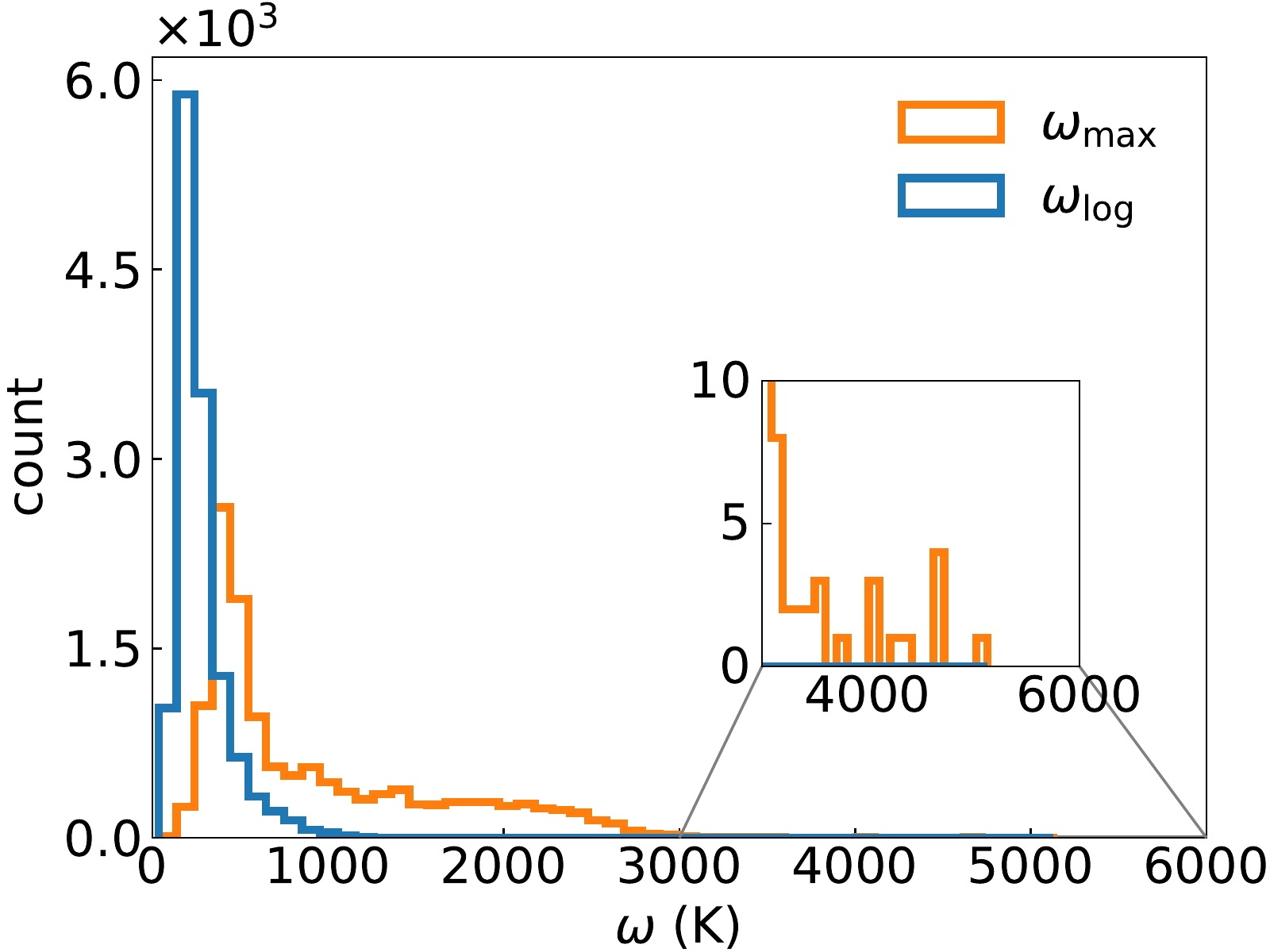}\\
    \includegraphics[width=0.9\columnwidth]{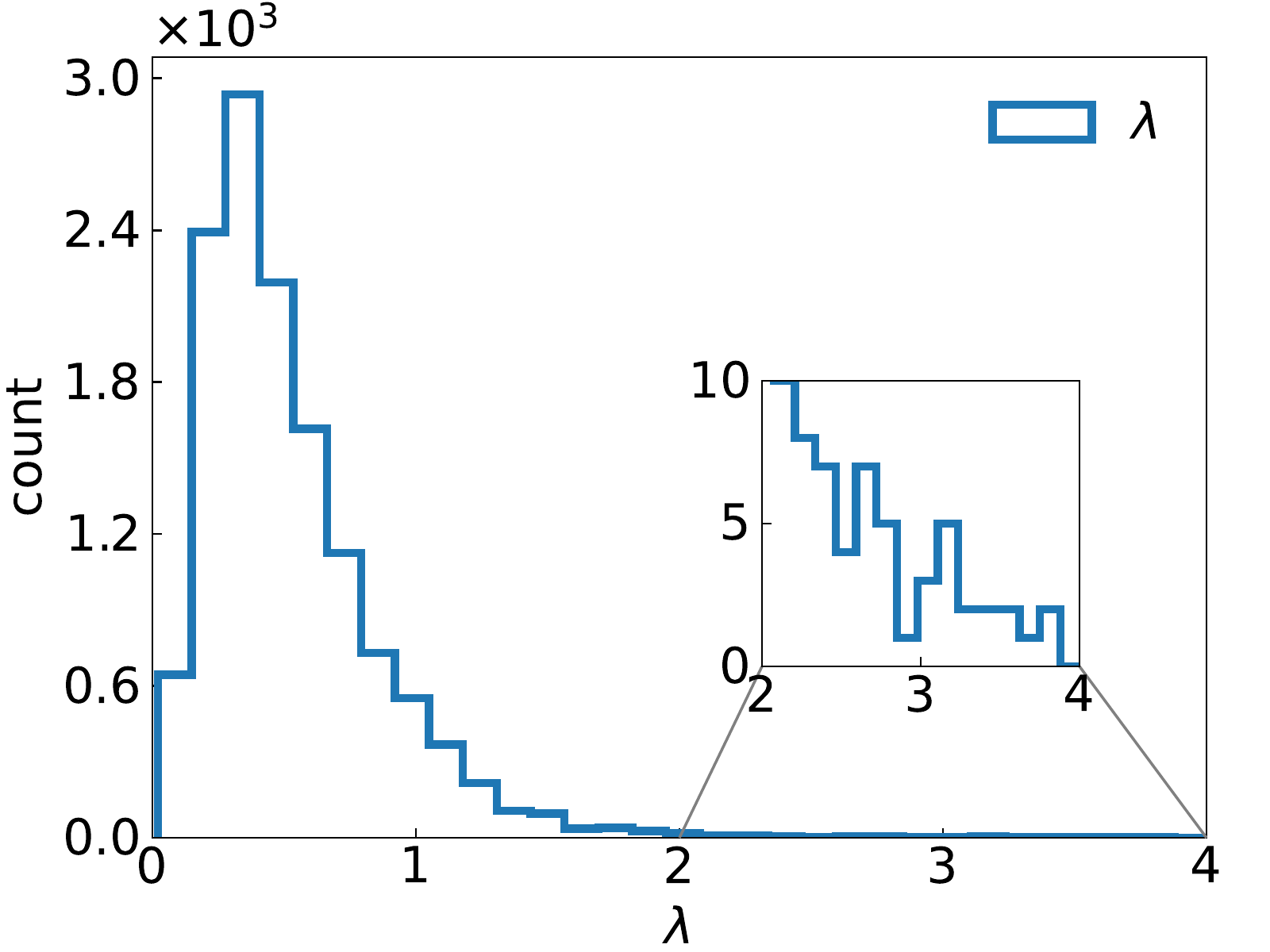}\\
    \caption{Histogram of (top) the maxima (\omax) and logarithmic average (\olog) of phonon frequencies and (bottom) the electron-phonon coupling constants (\la) calculated for around 20\,000 metals (see Refs.~\cite{cerqueira2024searching}).}
    \label{fig:wmax_lambda_distr}
\end{figure}

The first step is to determine what is the largest phonon frequency $\omax$ that one can reasonably expect in a realistic compound. Reference~\onlinecite{trachenko2024upper} derived the expression
\begin{equation}
    \omax =  \frac{E^\text{electronic}}{\sqrt{m}}
    ,
\end{equation}
where $E^\text{electronic}$ is a typical electronic energy, and $m$ is an atomic mass. By inserting a typical $E^\text{electronic} = 1$~Ry, and the proton mass $m = m^\text{proton}$ into the formula, they obtained $\omax = 3680$~K. Compared to the vibrational frequency of \ce{H2} that is almost 6000~K~\cite{Huber1979}, this value does not appear to be an overestimation.

The same conclusion can be reached by looking at the calculated values for our materials depicted in \cref{fig:wmax_lambda_distr}. The distribution of values of \omax\ exhibits two maxima, one at around 500~K and another around 2000~K. Although the overwhelming majority of the compounds have \omax\ below 3000~K, we find a few that extend to 5400~K, well beyond the estimation of Ref.~\onlinecite{trachenko2024upper}. Not surprisingly, all of these are hydrides. An example of one such material, with a calculated $\omax=5396$~K, is the hypothetical \ce{AgTl2H2}. This is a compound that contains isolated \ce{H2} molecules inside a \ce{AgTl2} framework (see Supplementary Information, SI). The H--H distance is 0.78~\AA, only slightly larger than the value of 0.76~\AA\ in solid hydrogen calculated in the same approximation. As expected by the extreme difference of masses, the phonon band structure splits into separated manifolds, with the lowest lying states with Tl and Ag character and the highest states coming from H. The \ce{H2} stretching mode has negligible dispersion, and can be found at around 5400~K. The Ag--Tl and lowest lying H-modes all couple very strongly to the electrons, leading to the large $\lambda=1.1$ (and a $\Tc\sim11.5$~K). As expected by the very high frequency, the highest lying phonon mode has a negligible contribution to $\lambda$, even if $\alpha^2F(\omega)$ exhibits a very high peak at that frequency.

In \cref{fig:wmax_lambda_distr} we also plot the distribution of the values of $\olog$. Contrary to $\omax$, $\olog$ has a single peak at very low frequency, and decays rapidly with frequency. At the end of the tail, at $\olog$ values in excess of 1800~K we find a few hydrides, such as the hypothetical perovskite \ce{NaNiH3}, and some ordered crystals of boron-doped sp$^3$ carbon (see SI). In the case of the hydride, only the high frequency H-modes have a significant coupling to the electrons, leading to a large logarithmic average but a very small value of $\lambda$ and consequently of $\Tc$. The boron-doped case takes advantage of the high-energy of the carbon modes (due to the very strong C--C sp$^3$ bond) and from the fact that phonon modes couple strongly to the electrons in a large energy range. As expected, also in this case $\lambda$ has moderate values, of the order of 0.5--0.6, leading to \Tc\ in the range of 10--30~K.

As the determinant factor for the calculation of \Tc\ is \olog\ and not \omax, from this discussion it seems much more reasonable to use values of the order of 1800~K, and not $\omax = 3680$~K, as in Ref.~\onlinecite{trachenko2024upper}.

The next step in the estimation of the maximum value of \Tc\ is the optimization of the shape of $\alpha^2F(\omega)$, assuming a maximum phonon frequency of \omax. A free optimization of this function would obviously lead to $\Tc=\infty$, as $\alpha^2F(\omega)$ is not constrained by any sum-rule. Therefore, Trachenko \textit{et al.} fixed $\lambda = 2$ \cite{trachenko2024upper}, the value that maximizes $\Tc$  according to McMillan's formula. From the considerations above, and from the lower panel of \cref{fig:wmax_lambda_distr}, $\lambda = 2$ seems to be a reasonable value, perfectly reachable in a variety of compounds. The hypothetical compounds we have in our dataset with highest values of $\lambda\approx3.3$ are \ce{ClB2C8} and \ce{Al2OsH7} (see SI). The former is a C clathrate p-doped with B and co-doped with endohedral Cl, with a $\olog=425$~K and a calculated $\Tc=55$~K. The latter, that exhibits a \ce{Al2Os} framework that includes a large quantity of hydrogen, has an \olog\ of almost 300~K, leading to $\Tc=38$~K.

The optimal shape of $\alpha^2F(\omega)$ obtained by Ref.~\onlinecite{trachenko2024upper} by optimizing the \Tc\ calculated from the Eliashberg equations is a narrow peak at \omax. Actually, the  same conclusion follows directly from McMillan's formula for $T_c$. For a fixed value of $\lambda$, the value of \Tc\ changes linearly with \olog. In turn, the maximum value of \olog\ is obtained when $\alpha^2F(\omega) \sim \delta(\omega-\omax)$, leading to $\olog=\omax$. If, for the sake of the argument, we insert $\lambda=2$, $\olog=1800$~K into the McMillan function, and assume a value of $\mu^*=0.1$, we obtain $\Tc=260$~K. This is smaller than the value of Ref.~\onlinecite{trachenko2024upper}, but still much larger than the current record of \ce{MgB2}.

\begin{figure}[tb]
    \centering
    \includegraphics[width=0.95\columnwidth]{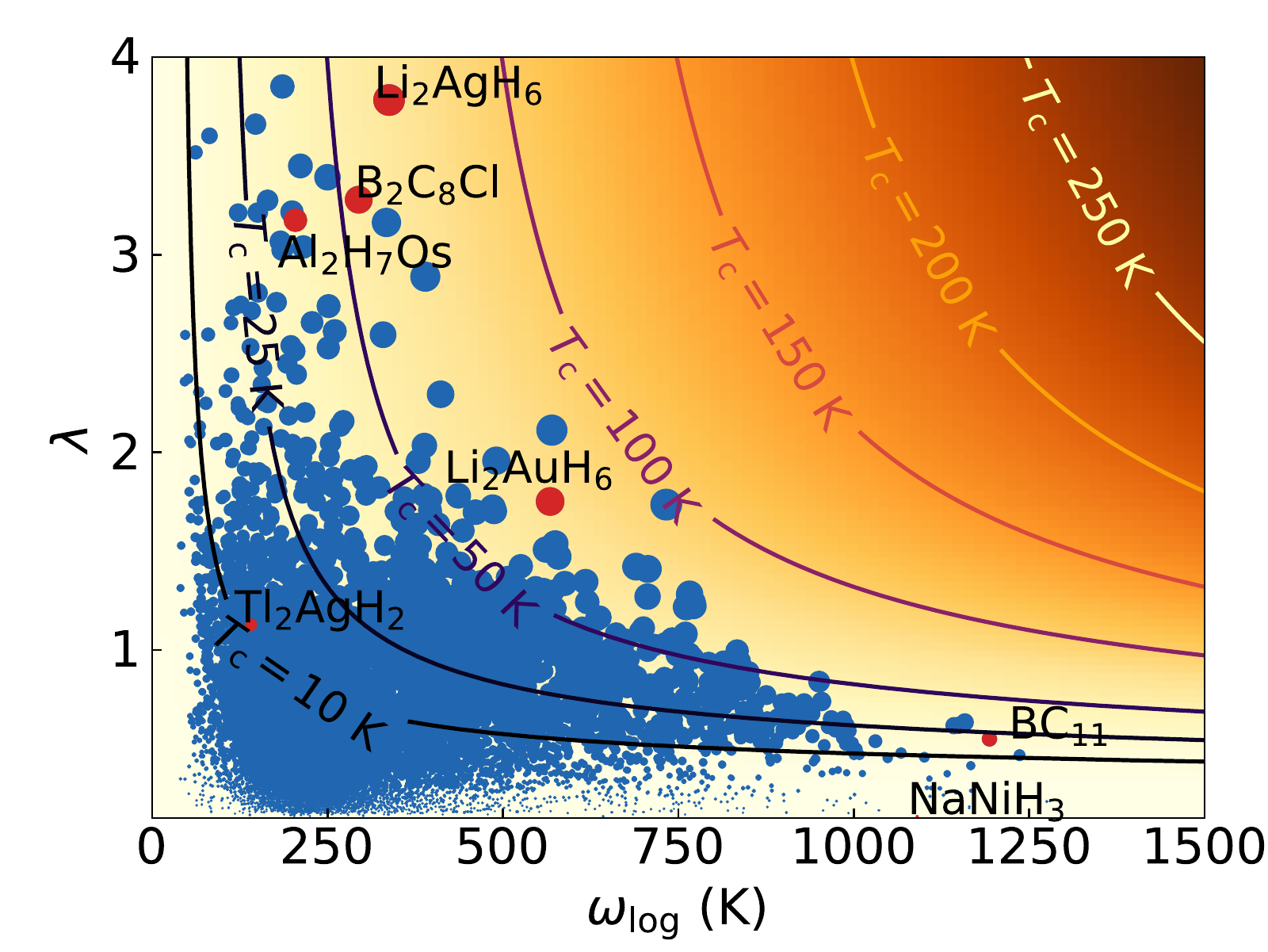}
    \caption{Scatter plot of \olog\ versus \la\ for all calculated systems, the contour lines are \Tc\ from McMillan's formula with $\mu^*=0.1$.}
    \label{fig:wlog_lambda}
\end{figure}

It is most likely, however, that this limit is unattainable for any physical system at ambient pressure. In fact, the parameters $\olog$ and $\lambda$ are not entirely independent, as they are two different moments of the same $\alpha^2F(\omega)$ function. This was already recognized in the seminal work of McMillan in 1968~\cite{mcmillan1968transition}, where it was recognized that the most important dependence of $\lambda$ is om an average phonon frequency. Obviously \olog\ is favored by high frequencies, while $\lambda$ by low frequencies, so the shape of the optimal $\alpha^2F(\omega)$ that maximizes \Tc\ (and that only contains a \textit{single}, very high-frequency, flat-band phonon mode that couples very strongly to the electrons) seems to be unattainable physically.

This evidence is also supported by the data depicted in \cref{fig:wlog_lambda}, where we plot the relationship between \olog\ and $\lambda$ for all compounds in our dataset. The size of the circles is proportional to \Tc\ and in the background we plot the contour lines of constant \Tc\ as calculated from McMillan's formula with $\mu^*=0.1$. We see that, as expected, compounds with very large values of \olog\ often have small $\lambda$ and vice-versa. Furthermore, no material comes close to the optimal case, and compounds with the highest values of \Tc\ are the ones that achieve a good compromise between reasonable high \olog\ and $\lambda$.

\begin{figure}[tb]
    \centering
    \includegraphics[width=0.99\columnwidth]{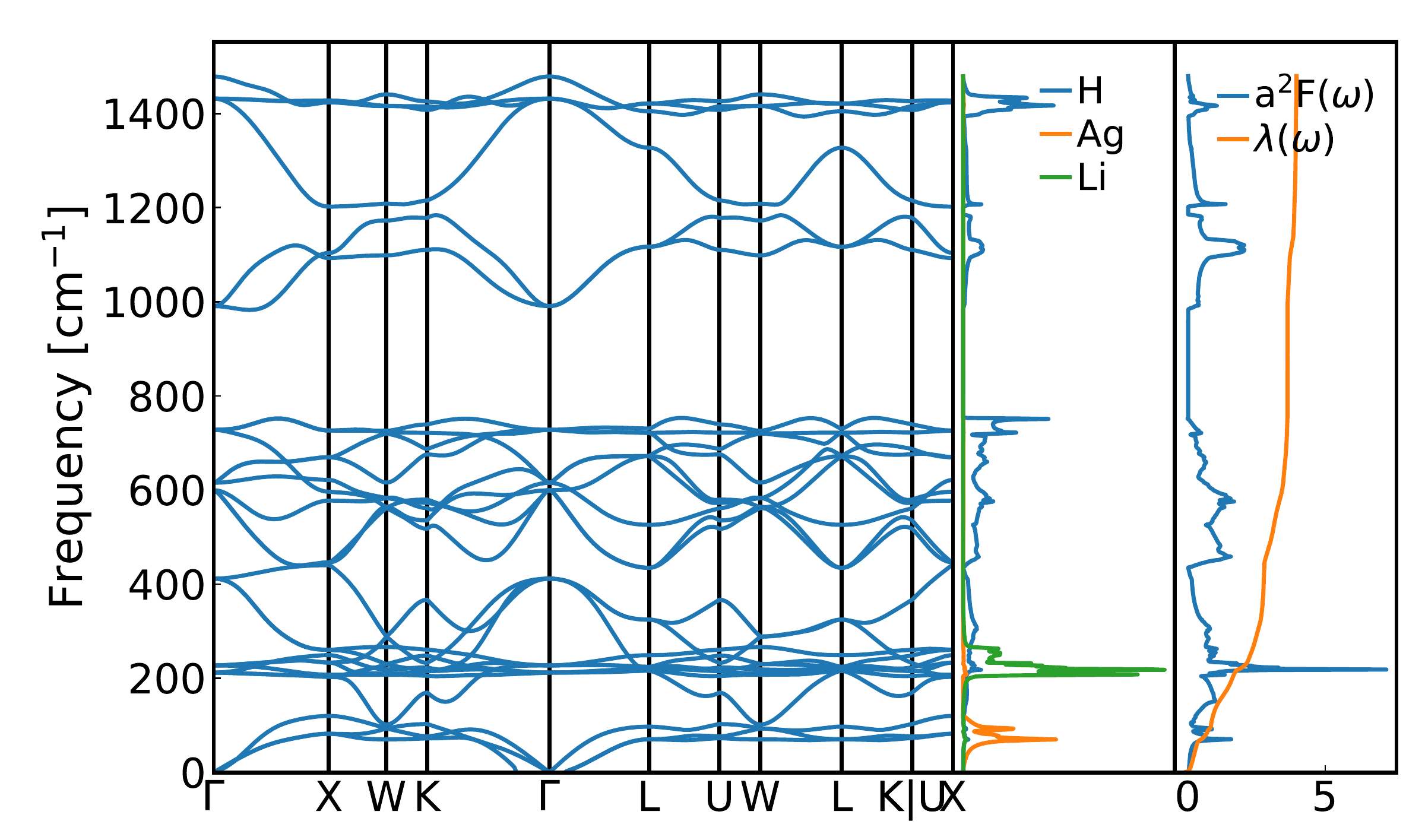}
       \caption{Harmonic phonon band structure, density of phonon states, and Eliashberg spectral function $\alpha^2F(\omega)$ of \ce{Li2AgH6} calculated within the harmonic approximation. }
     \label{fig:hightccompound}
\end{figure}

As an example, we will look at the compounds with the highest \Tc\ in our dataset, specifically \ce{Li2AgH6} and \ce{Li2AuH6} (see \cref{fig:hightccompound} and SI). These compound crystallize in the same cubic structure as  \ce{Mg2IrH6}, \ce{Mg2PdH6}, \ce{Mg2PtH6}, etc. that were recently proposed~\cite{sanna2024prediction,hansen2024synthesis} as high-\Tc\ superconductors. 
Both materials are thermodynamically unstable at ambient pressure at respectively 0.319~eV/atom and 0.172~eV/atom above the convex hull for the Ag and the Au compounds, and are not significantly stabilized by pressure (at least up to 50~GPa, see SI).

The electronic band structure of \ce{Li2AgH6} resembles the one of \ce{Mg2IrH6}, with a single band crossing the Fermi surface. This band, with very strong H-character, is however more dispersive in the present case. Also the phonon band structure is similar, with the acoustic modes mostly composed by vibrations of the heavier atom, followed by modes stemming from the alkali or alkaline earth metal. Finally, there are three separate manifolds of phonon bands. In \ce{Li2AgH6} the lowest manifold contains both H and Li character due to the small difference between the atomic masses of these elements. Essentially all modes contribute to the very high value of $\lambda\approx4$.

We would like to note that one of the acoustic phonon branches exhibits (small) imaginary values close to the $\Gamma$. This is a shortcoming of the harmonic approximation, and the structure is perfectly dynamically stable when anharmonic and quantum nuclear effects are taken into account (see SI)~\cite{Errea-PRB2014-SSCHA,Monacelli-JPCM2021-SSCHA}. In any case, 95\% of the electron-phonon coupling comes from mid-low frequency modes which are almost identical between harmonic and anharmonic calculations.

In order to obtain a more accurate \Tc\ estimation within modern superconductivity methods we have recomputed the value of \Tc\ using two state-of-the art approaches which, unlike more conventional methods, include the electron phonon coupling and the Coulomb interaction calculated from first-principles. These are the full Eliashberg approach of Ref.~\onlinecite{Pellegrini_SimplifiedEliashberg2022} and density functional theory for superconductors~\cite{Pellegrini_NatRev2024,Sanna_PRL_2020_combining,OGK_PRL_1988,Lueders_SCDFTI,Marques_SCDFTII} (SCDFT). For the sake of these high-accuracy calculations the electron-phonon coupling was recomputed using a Monte-Carlo $k$-mesh accumulated on the Fermi surface, on which the electron-phonon matrix elements are linearly interpolated~\cite{Sanna_anisotropyNbSe2_2022}. This ensures a perfect convergence of the nesting properties entering the definition of the Eliashberg spectral function.

The resulting critical temperature for \ce{Li2AhH6} is 108.8~K in Eliashberg theory and 83.0~K in SCDFT. Although both approaches are derived from the Migdal approximation for the electron-phonon self energy~\cite{Migdal_1958} and assume a static Coulomb interaction, the slightly different \Tc\ prediction is dictated by the approximation chain that leads to each computational scheme. On one hand, in SCDFT there is the approximation to the anomalous exchange-correlation functional~\cite{Sanna_PRL_2020_combining}, which adds additional approximations to the form of the self-energy. On the other hand, in Eliashberg theory the energy dependence of the Coulomb interaction is neglected at the scale of the phononic energies.  However, considering that the electron-phonon coupling parameter $\lambda$ is very high, the functional approximation of SCDFT might be slightly beyond its validity range~\cite{Sanna_PRL_2020_combining}. Therefore we expect that, in this case, the Eliashberg estimation should be more accurate.

It is interesting that the Eliashberg estimation of \Tc\ is in very good agreement with the one obtained by means of the McMillan-Allen-Dynes approach using a standard $\mu^*=0.1$. This indicates that Coulomb interactions act as expected for a conventional sp-system~\cite{AllenMitrovic1983}. This differs from what is observed in \ce{Mg2IrH6} which, as discussed in Ref.~\onlinecite{sanna2024prediction}, has its \Tc\ overestimated by the $\mu^*$ model~\cite{MorelAnderson_mustar_PR1962}. The reason is the presence of a peak in the density of states at the Fermi level, that is close to a large band gap, leading to a poor Coulomb renormalization~\cite{AllenMitrovic1983,ScalapinoSchriefferWilkins_PRB1966}. The \ce{Li2AgH6} system also features a peak in the density-of-states and a band gap, however the peak is broader, while the band gap is very small, leading to an overall smoother density profile and efficient Coulomb renormalization.

A similar analysis can be extended to \ce{Li2AuH6}. This system has electronic, phononic and superconducting properties nearly identical to its Ag twin. The predicted superconducting \Tc\ including Coulomb interactions is 91.0~K and 116.1~K in SCDFT and Eliashberg, respectively.

In view of the discussion before, these materials seem to be ideal cases for conventional superconductivity, and their \Tc\ is likely in the maximum range of what can be achievable at ambient pressure. We note that these are
isotropic superconductors, where the effect of anisotropy accounts for less than 1\% of the value of \Tc. This is the opposite of \ce{MgB2}, where the electron-phonon coupling mostly acts on the $\sigma$ bands, and an isotropic calculations leads to an underestimation of \Tc\ by almost a factor of two. The isotropic superconducting state of \ce{Li2AgH6} or \ce{Li2AuH6}, if experimentally realized, would not only have a critical temperature above liquid nitrogen, but it would also be more suitable for high-field applications. In fact these applications require impurities and defects to work as pinning centers for magnetic flux lines and reduce the Ginsburg-Landau coherence-length. Isotropic superconductors are more likely to be stable upon the introduction of crystalline defects and still be affected by their scattering potential.

\begin{figure}[tb]
    \centering
    \includegraphics[width=0.92\columnwidth]{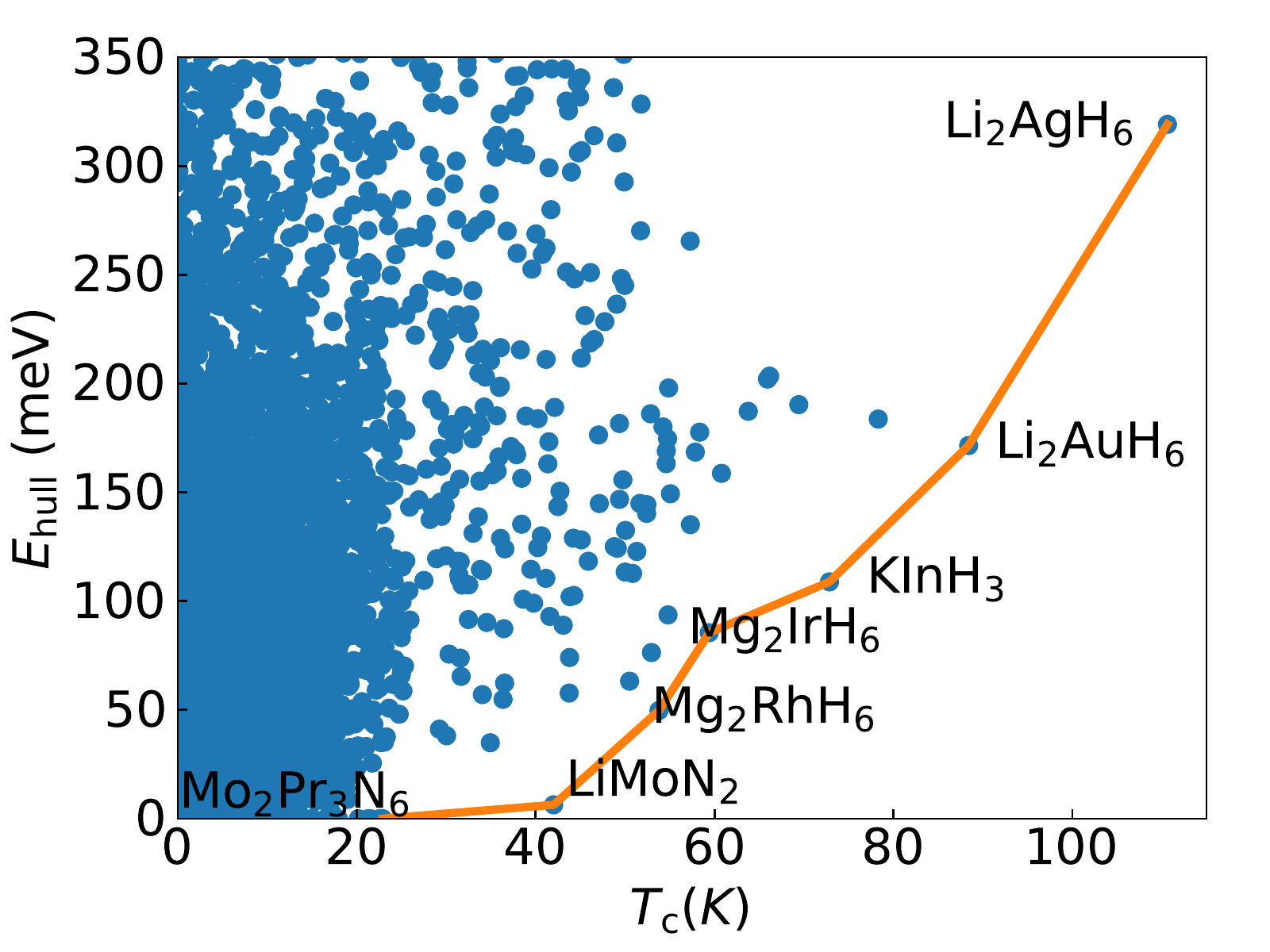}
    \caption{Scatter plot of the distance to the convex hull of stability versus the superconducting transition temperature for all compounds in our dataset. In orange we also indicate the Pareto front corresponding to this data. Compounds on the Pareto front are labeled.}
    \label{fig:Tc_ehull}
\end{figure}

Finally, we would like to discuss the issue of synthesizeability. Most of the compounds discussed above, and others that have been proposed in the literature, with high-\Tc\ are hypothetical, and up to now none has been stabilized at ambient pressure. To better understand this problem, we plot in \cref{fig:Tc_ehull} the distance to the convex hull of thermodynamic stability and the calculated transition temperature of all compounds in our dataset. We also indicate the Pareto front that corresponds to this data. The values of \Tc\ are obtained with an isotropic theory, so the transition temperature of \ce{MgB2} is considerably underestimated. We also note that although being close to the hull is not a synonym of synthesizeability, the higher the distance to the hull for a given compound, the smaller the probability that this compound can be stabilized experimentally.

Very close to the hull, we find that the compound with highest predicted \Tc\ is \ce{LiMoN2} with around 40~K~\cite{cerqueira2024sampling}. Unfortunately, this compound exhibits intrinsic defects that lower the high density of states at the Fermi level and destroy superconductivity~\cite{Elder1992, cerqueira2024sampling}. If it is possible to resolve this problem experimentally and synthesize the pristine compound is at this point unknown~\cite{hunter2008structural}. At higher values of \Tc\ we find several hydrides and boron-carbides. Unfortunately, all these compounds are unstable thermodynamically with a distance to the hull that increases rapidly along the Pareto front. This is easy to understand as hydrides prefer charge-compensated, semiconducting (or insulating) phases and are therefore destabilized in the metallic phase, and boron induces stresses in the very strong diamond framework increasing its energy. A possible path through the high-pressure synthesis followed by quenching of these compounds to ambient pressure is commonly suggested in the literature ~\cite{Zurek_ChemicalPrecompression_JAP2022,Lucrezi_npj2022,Dangic-arxiv2024-RbPH3}, however it is still unproved to this date, and remains highly speculative.

In summary, by analyzing data from more than 20\,000 electron-phonon calculations we critically discussed the possible maximum \Tc\ of conventional superconductors at ambient pressure. It seems clear that it is possible to design hypothetical compound with values of \Tc\ reaching 100--120~K, much larger than the current experimental record, but still very far from room temperature. Unfortunately, all compounds with high \Tc\ appear to be thermodynamically unstable, raising questions about their experimental synthesis and characterization. It is true that physical laws do not restrict \Tc\ to go beyond 100--120~K, but in practice our data show that the experimental realization of a compound with such high \Tc\ is extremely unlikely.

\section*{Acknowledgements}
This work was supported by Simons Foundation through the Collaboration on New Frontiers in Superconductivity (Grant No. SFI-MPS-NFS-00006741-10) and by the Keele foundation.
T.F.T.C acknowledges financial support from FCT - Fundação para a Ciência e Tecnologia, I.P. through the project CEECINST/00152/2018/CP1570/CT0006 with
DOI identifier 10.54499/CEECINST/00152/2018/CP1570/\linebreak{}CT0006, and computing resources provided by the project Advanced Computing Project 2023.14294.CPCA.A3, platform Deucalion. H.-C.W and M.A.L.M acknowledge the computing time on the high-performance computer Noctua 2 at the NHR Center PC2. Y.-W.F., {\DJ}.D., and I.E. acknowledge funding from the European Research Council (ERC) under the European Union’s Horizon 2020 research and innovation program (grant agreement No. 802533) as well as EuroHPC for granting the access to Lumi located in CSC’s data center in Kajaani, Finland, (Project ID EHPC-REG-2024R01-084) and the technical and human support provided by DIPC Supercomputing Center. I.E. also received funding the Spanish Ministry of Science and Innovation (Grant No. PID2022-142861NA-I00), and the Department of Education, Universities and Research of the Eusko Jaurlaritza and the University of the Basque Country UPV/EHU (Grant No. IT1527-22). K.G. acknowledges financial support from the China Scholarship Council.
The authors acknowledge enlightening discussions with the partners of the SuperC collaboration.

\bibliography{references.bib}

\clearpage
\newpage
\begin{figure*}
\includegraphics[page = 1, width=\linewidth]{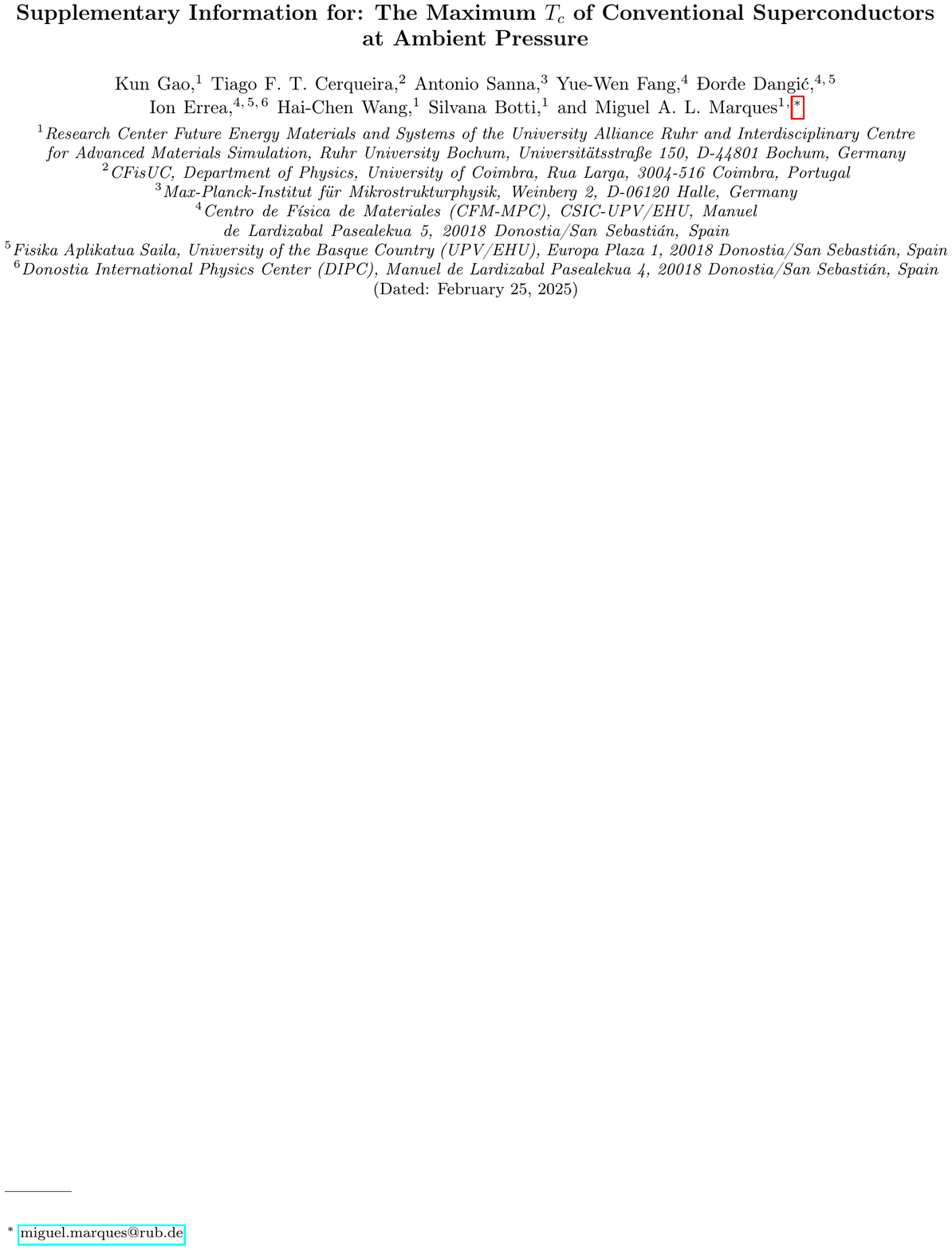}
\end{figure*}
\begin{figure*}
\includegraphics[page = 2, width=\linewidth]{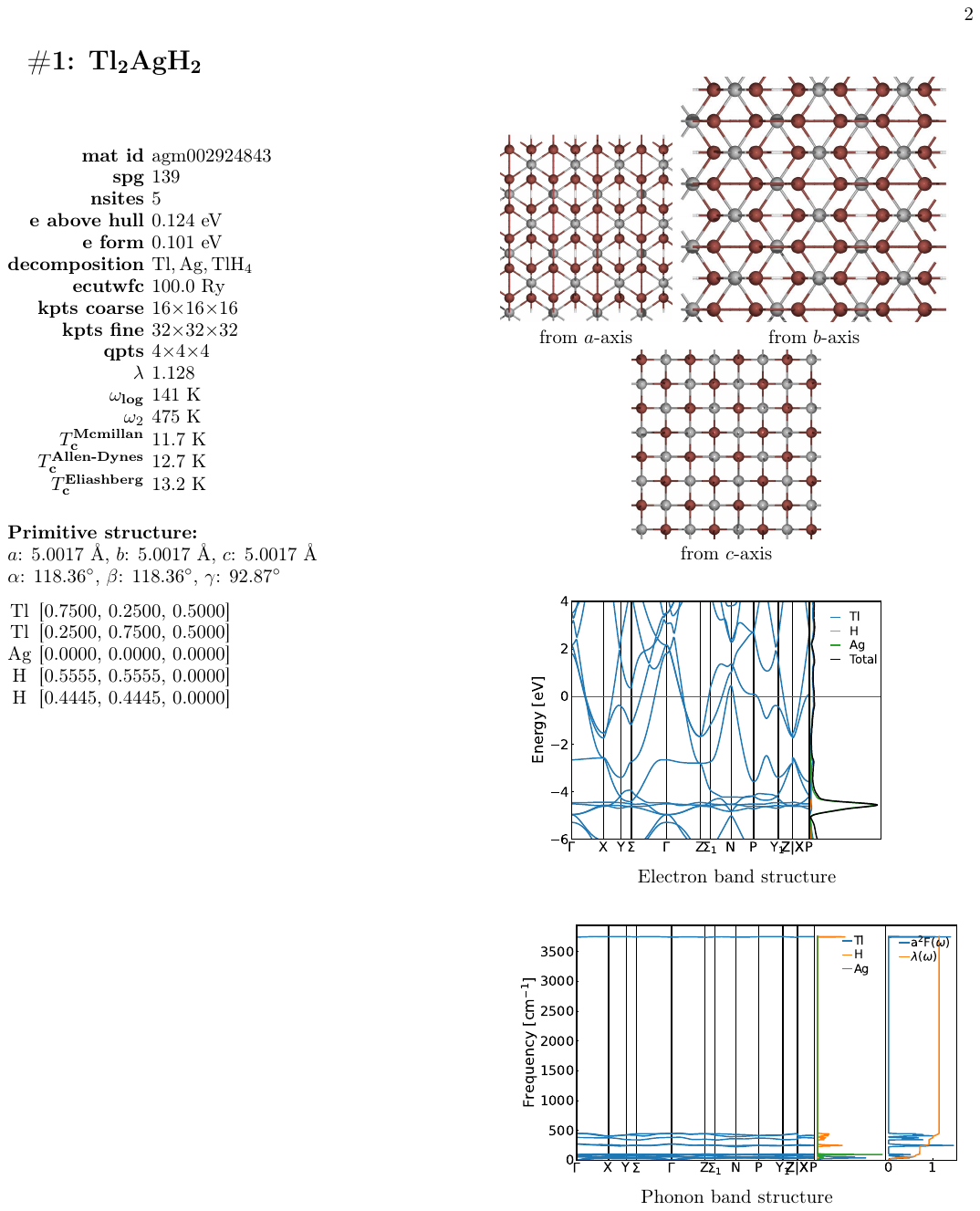}
\end{figure*}
\begin{figure*}
\includegraphics[page = 3, width=\linewidth]{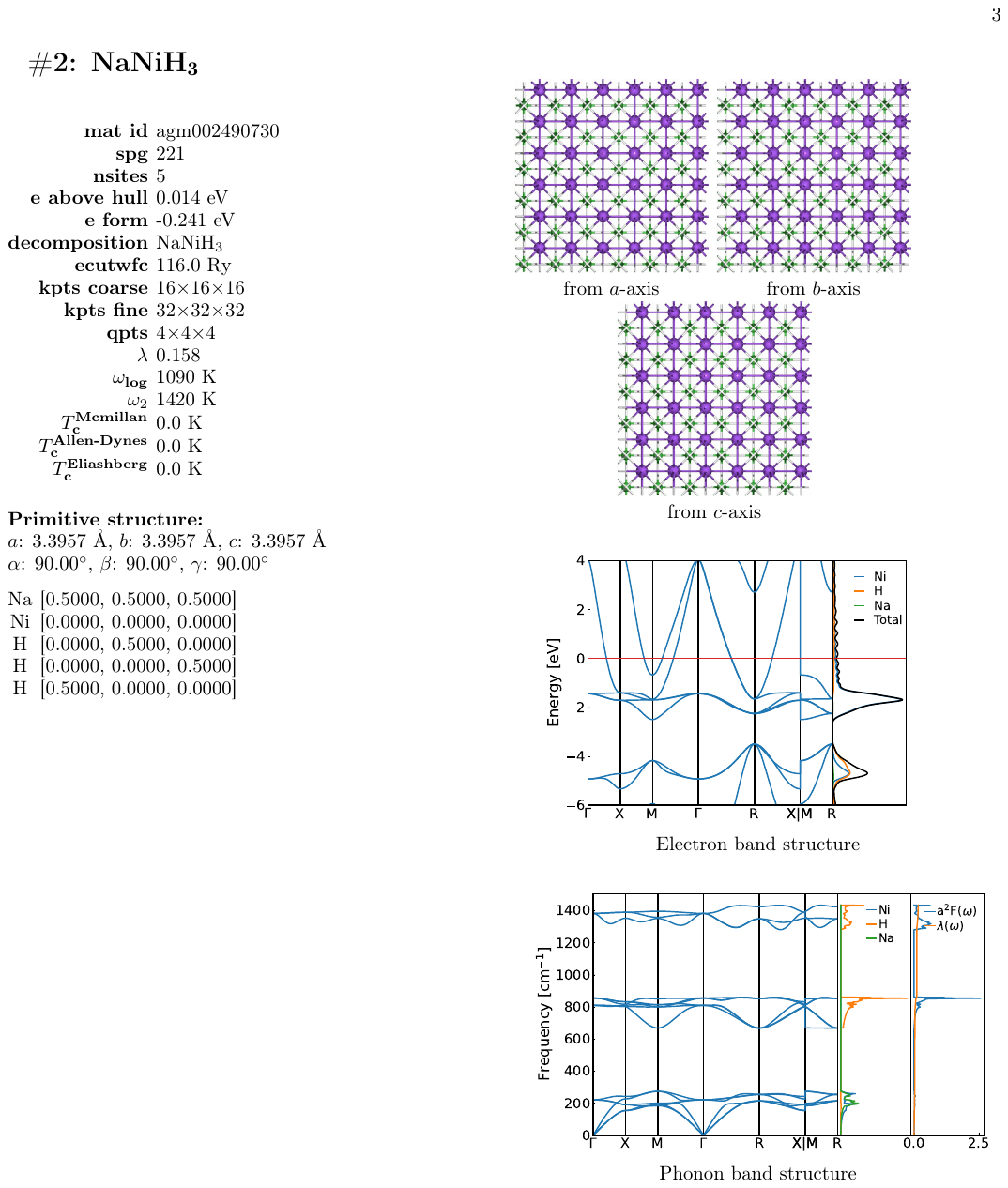}
\end{figure*}
\begin{figure*}
\includegraphics[page = 4, width=\linewidth]{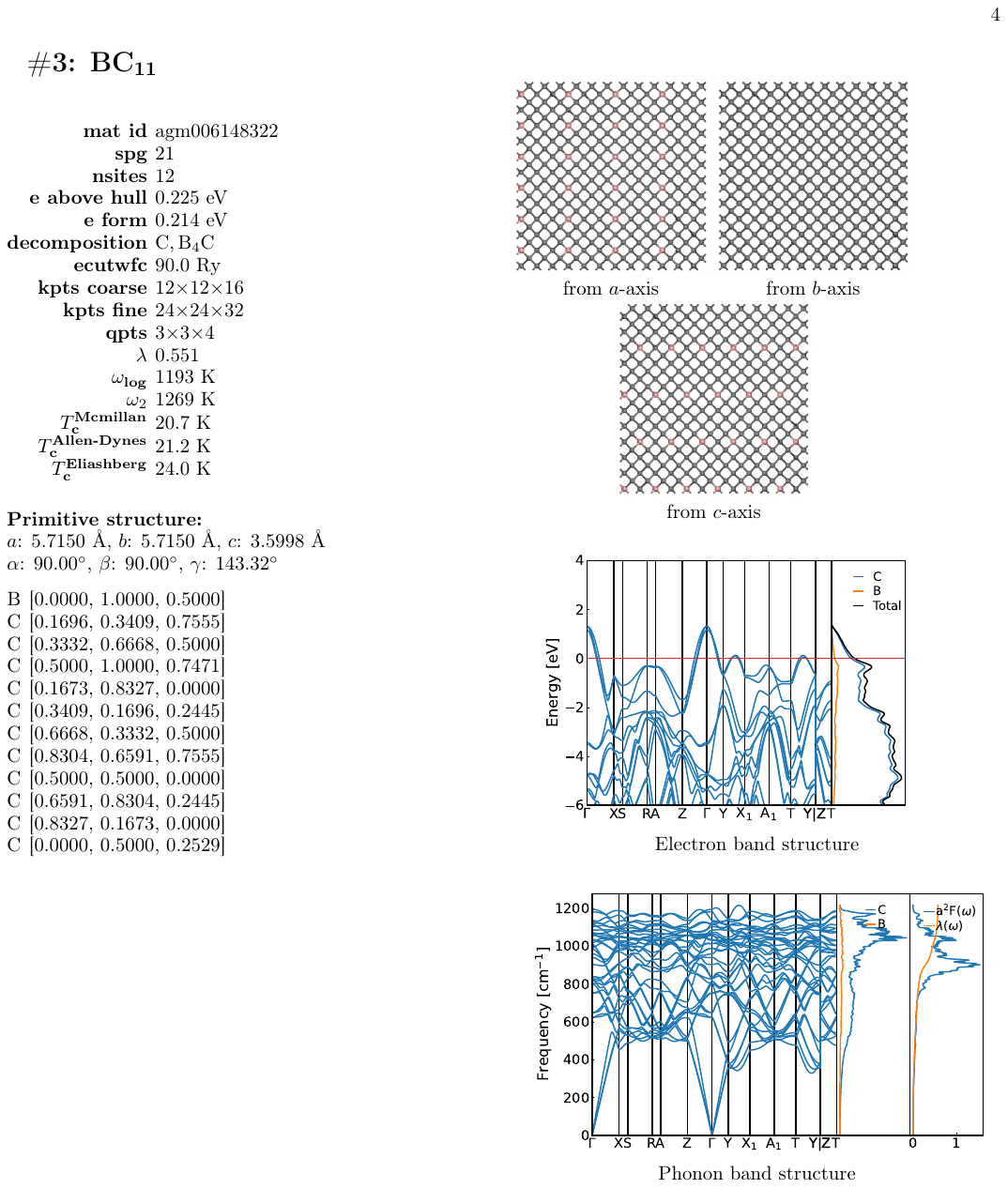}
\end{figure*}
\begin{figure*}
\includegraphics[page = 5, width=\linewidth]{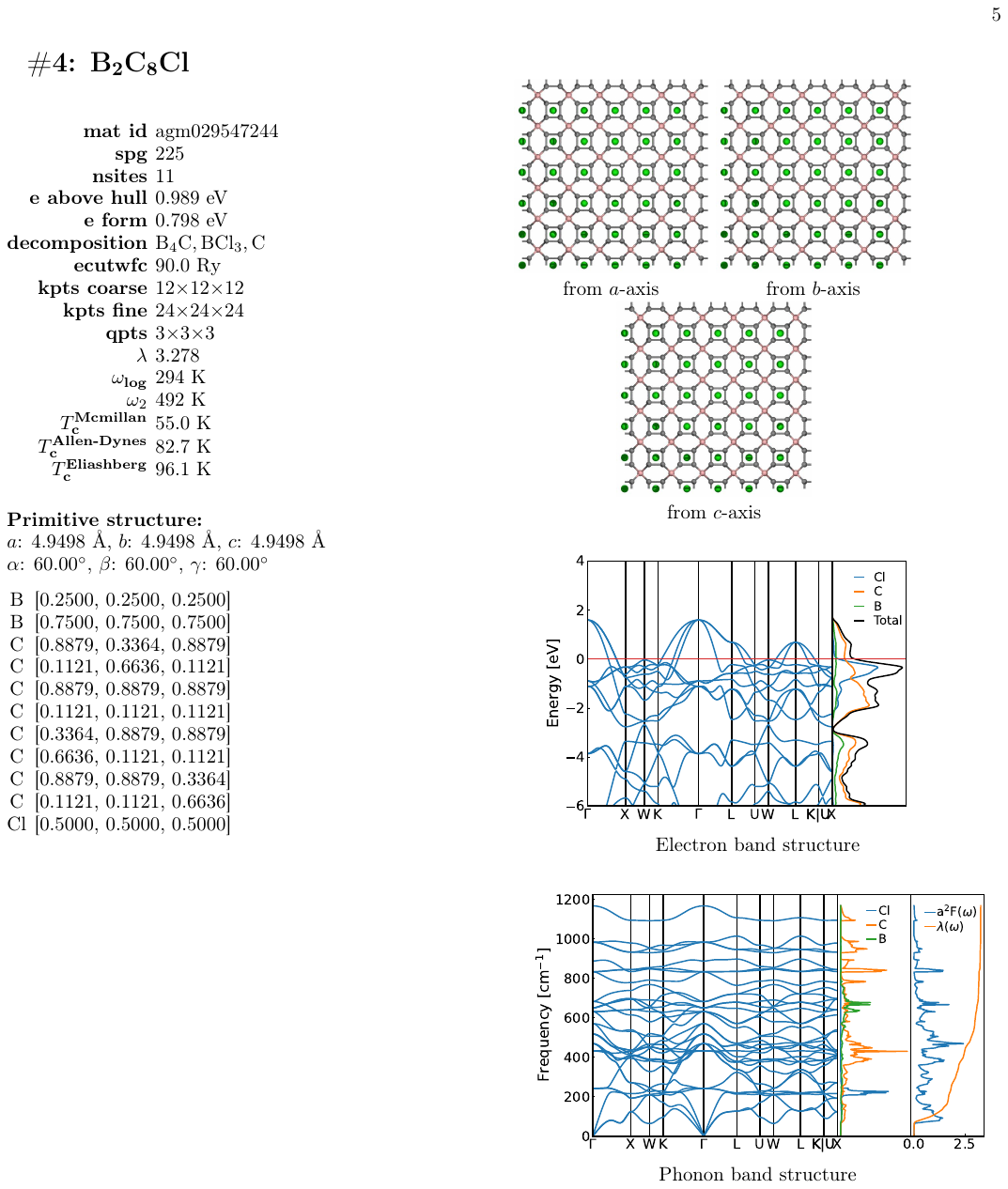}
\end{figure*}
\begin{figure*}
\includegraphics[page = 6, width=\linewidth]{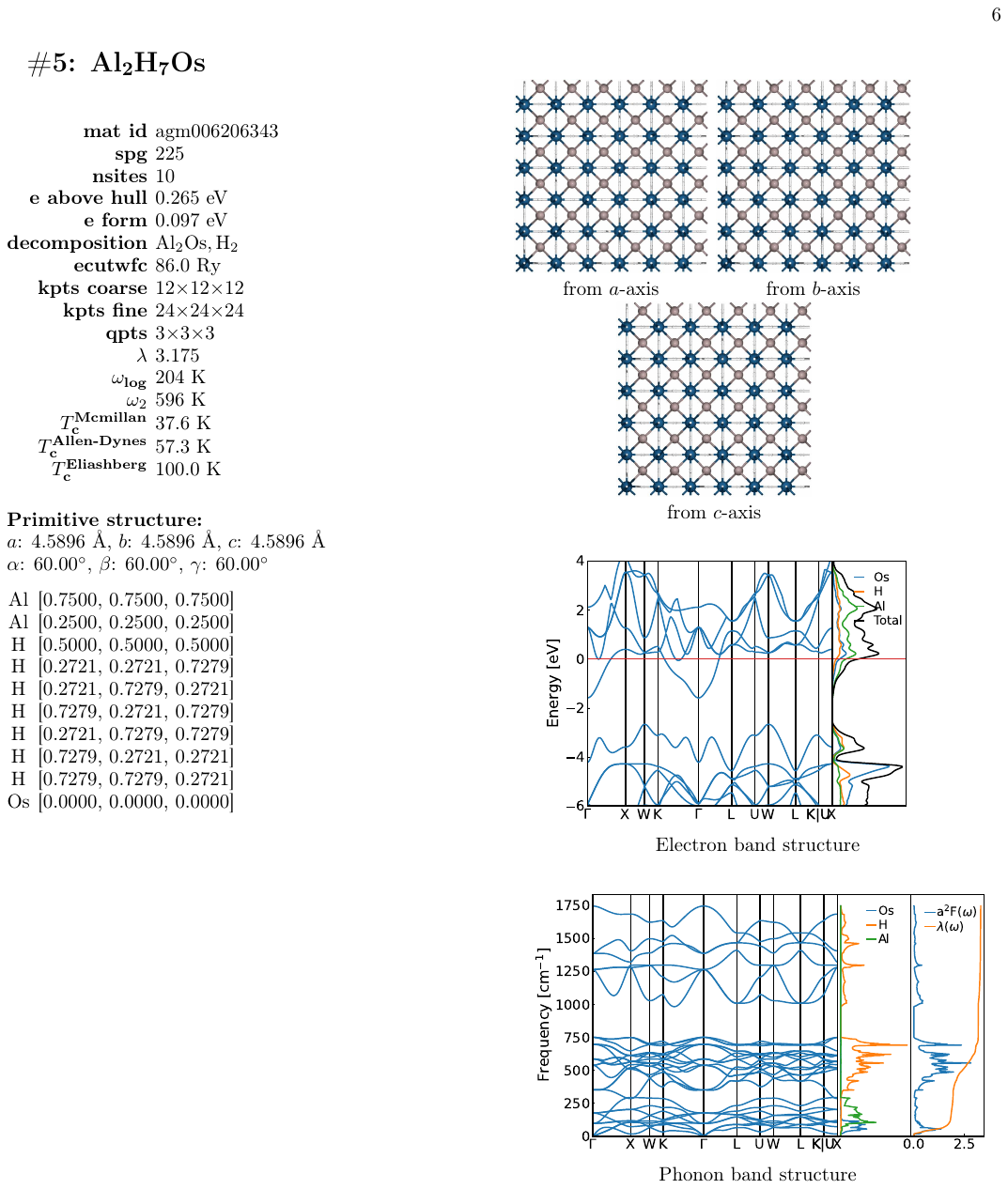}
\end{figure*}
\begin{figure*}
\includegraphics[page = 7, width=\linewidth]{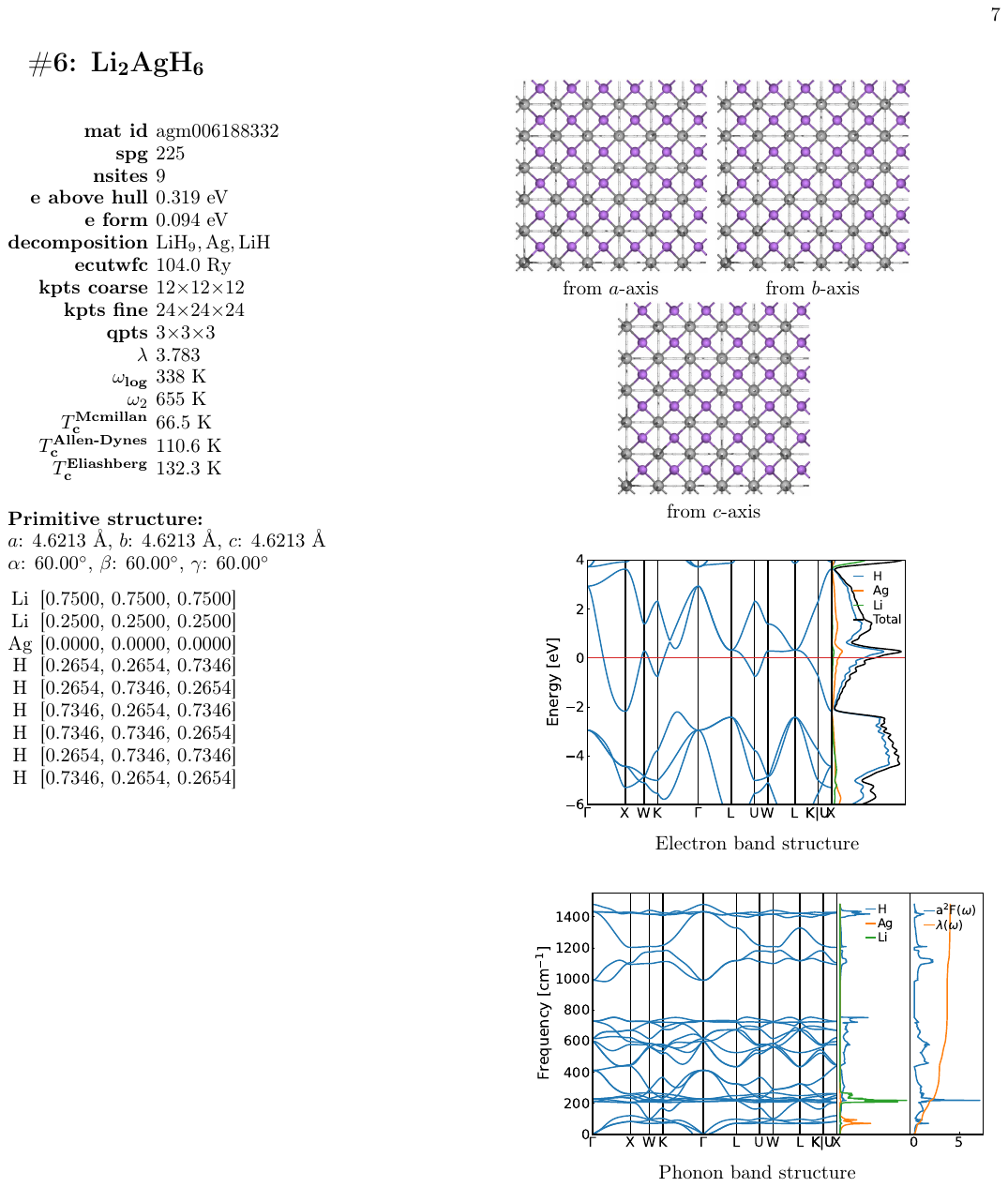}
\end{figure*}
\begin{figure*}
\includegraphics[page = 8, width=\linewidth]{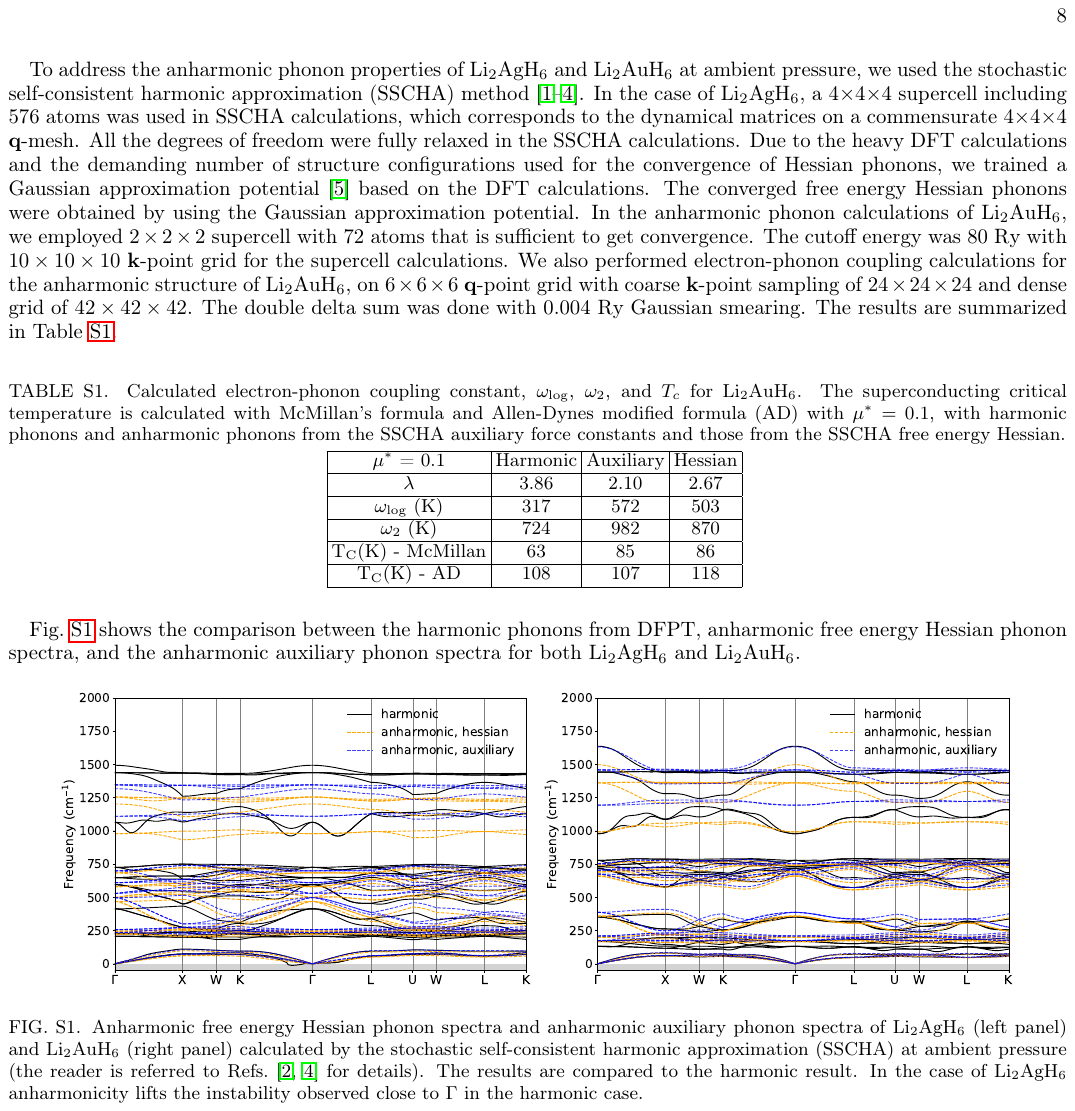}
\end{figure*}
\begin{figure*}
\includegraphics[page = 9, width=\linewidth]{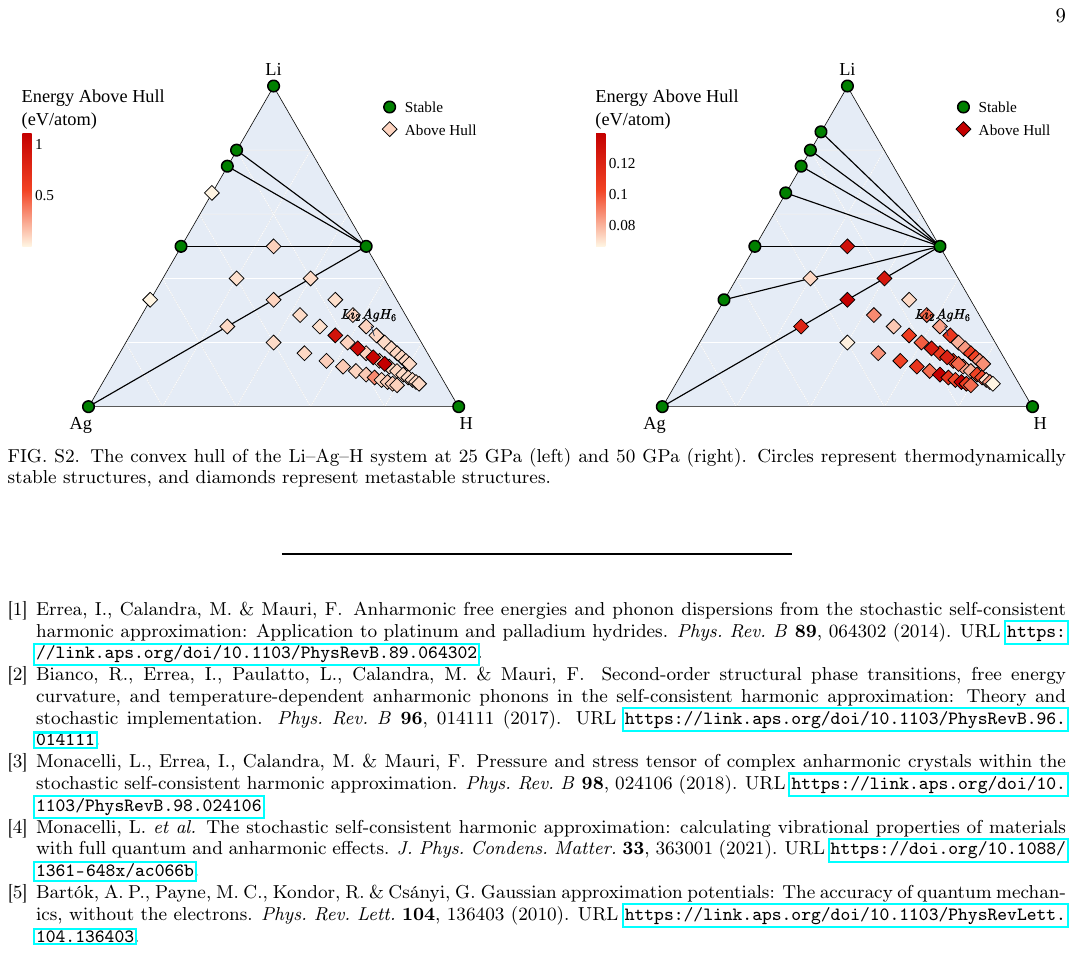}
\end{figure*}
\end{document}


\newcommand{\bochum}{Research Center Future Energy Materials and Systems of the University Alliance Ruhr and Interdisciplinary Centre for Advanced Materials Simulation, Ruhr University Bochum, Universit\"atsstraße 150, D-44801 Bochum, Germany}
\newcommand{\coimbra}{CFisUC, Department of Physics, University of Coimbra, Rua Larga, 3004-516 Coimbra, Portugal}
\newcommand{\mpihalle}{Max-Planck-Institut f\"ur Mikrostrukturphysik, Weinberg 2, D-06120 Halle, Germany}
\newcommand{\ehu}{Fisika Aplikatua Saila, University of the Basque Country (UPV/EHU), Europa Plaza 1, 20018 Donostia/San Sebasti{\'a}n, Spain}
\newcommand{\cfm}{Centro de F{\'i}sica de Materiales (CFM-MPC), CSIC-UPV/EHU, Manuel de Lardizabal Pasealekua 5, 20018 Donostia/San Sebasti{\'a}n, Spain}
\newcommand{\dipc}{Donostia International Physics Center (DIPC), Manuel de Lardizabal Pasealekua 4, 20018 Donostia/San Sebasti{\'a}n, Spain}

\author{Kun Gao}
\affiliation{\bochum} 
\author{Tiago F. T. Cerqueira}
\affiliation{\coimbra}
\author{Antonio Sanna}
\affiliation{\mpihalle}
\author{Yue-Wen Fang}
\affiliation{\cfm}
\author{{\DJ}or{\dj}e Dangi{\'c}}
\affiliation{\cfm}
\affiliation{\ehu}
\author{Ion Errea}
\affiliation{\cfm}
\affiliation{\ehu}
\affiliation{\dipc}
\author{Hai-Chen Wang}
\affiliation{\bochum} 
\author{Silvana Botti}
\affiliation{\bochum} 
\author{Miguel A. L. Marques} 
\email{miguel.marques@rub.de}
\affiliation{\bochum} 

\date{\today}

\newcommand{\Tc}{\ensuremath{T_\textrm{c}}}
\newcommand{\Tcmac}{\ensuremath{T_\textrm{c}^\text{McMillan}}}
\newcommand{\Tcad}{\ensuremath{T_\textrm{c}^\text{Allen-Dynes}}}
\newcommand{\Tce}{\ensuremath{T_\textrm{c}^\text{Eliashberg}}}
\newcommand{\olog}{\ensuremath{\omega_\text{log}}}
\newcommand{\omax}{\ensuremath{\omega_\text{max}}}
\newcommand{\dosef}{\ensuremath{\text{DOS}({\text{E}_\text{F}}})}
\newcommand{\afo}{\ensuremath{\alpha^2F(\omega)}}
\newcommand{\la}{\ensuremath{\lambda}}

\renewcommand{\thefigure}{S\arabic{figure}}
\renewcommand{\thetable}{S\arabic{table}}

\title{Supplementary Information for: The Maximum $T_c$ of Conventional Superconductors at Ambient Pressure}

\maketitle


\newpage

\newlength\Colsep
\setlength\Colsep{10pt}
\include{high_tc_latex/agm002924843}
\include{high_tc_latex/agm002490730}
\include{high_tc_latex/agm006148322}
\include{high_tc_latex/agm029547244}
\include{high_tc_latex/agm006206343}
\include{high_tc_latex/agm006188332}

To address the anharmonic phonon properties of \ce{Li2AgH6} and \ce{Li2AuH6} at ambient pressure, we used the stochastic self-consistent harmonic approximation (SSCHA) method~\cite{Errea-PRB2014-SSCHA,Bianco-PRB2017-SSCHA,monacelli2018pressure,Monacelli-JPCM2021-SSCHA}. In the case of \ce{Li2AgH6}, a 4$\times$4$\times$4 supercell including 576 atoms was used in SSCHA calculations, which corresponds to the dynamical matrices on a commensurate 4$\times$4$\times$4 \textbf{q}-mesh. 
All the degrees of freedom were fully relaxed in the SSCHA calculations. Due to the heavy DFT calculations and the demanding number of structure configurations used for the convergence of Hessian phonons, we trained a Gaussian approximation potential~\cite{PhysRevLett.104.136403} based on the DFT calculations. The converged free energy Hessian phonons were obtained by using the  Gaussian approximation potential.  In the anharmonic phonon calculations of \ce{Li2AuH6}, we employed $2\times 2\times 2$ supercell with 72 atoms that is sufficient to get convergence. The cutoff energy was 80 Ry with $10\times 10\times 10$ $\mathbf{k}$-point grid for the supercell calculations. We also performed electron-phonon coupling calculations for the anharmonic structure of \ce{Li2AuH6}, on $6\times 6\times 6$ $\mathbf{q}$-point grid with coarse $\mathbf{k}$-point sampling of $24\times 24\times 24$ and dense grid of $42\times 42\times 42$. The double delta sum was done with 0.004 Ry Gaussian smearing. The results are summarized in Table \ref{tab:li2auh6}.

\begin{table}[tbh!]
\caption{Calculated electron-phonon coupling constant, $\omega _\textrm{log}$, $\omega _2$, and $T_c$ for \ce{Li2AuH6}. The superconducting critical temperature is calculated with McMillan's formula and Allen-Dynes modified formula (AD) with $\mu^*=0.1$, with harmonic phonons and anharmonic phonons from the SSCHA auxiliary force constants and those from the SSCHA free energy Hessian.}
\begin{tabular}{|c|c|c|c|}
\hline
$\mu^*$ = 0.1                  & Harmonic & Auxiliary & Hessian \\ \hline
$\lambda$                      & 3.86 & 2.10      & 2.67    \\ \hline
$\omega _\textrm{log}$ (K)     & 317  & 572       & 503     \\ \hline
$\omega _2$ (K)                & 724  & 982       & 870     \\ \hline
T$_\textrm{C}$(K) - McMillan   & 63   & 85        & 86      \\ \hline
T$_\textrm{C}$(K) - AD         & 108  & 107       & 118     \\ \hline
\end{tabular}
\label{tab:li2auh6}
\end{table}

Fig.~\ref{sscha-phonon} shows the comparison between the harmonic phonons from DFPT, anharmonic free energy Hessian phonon spectra, and the anharmonic auxiliary phonon spectra for both \ce{Li2AgH6} and \ce{Li2AuH6}.

\begin{figure}[tbh!]
    \centering
    \includegraphics[width=0.45\columnwidth]{img/SSCHA-phonon-Li2AgH6.pdf}
    \includegraphics[width=0.45\columnwidth]{img/SSCHA-phonon-Li2AuH6.pdf}
    \caption{Anharmonic free energy Hessian phonon spectra and anharmonic auxiliary phonon spectra of \ce{Li2AgH6} (left panel) and \ce{Li2AuH6} (right panel) calculated by the stochastic self-consistent harmonic approximation (SSCHA) at ambient pressure (the reader is referred to Refs.~\cite{Bianco-PRB2017-SSCHA,Monacelli-JPCM2021-SSCHA} for details). The results are compared to the harmonic result. In the case of \ce{Li2AgH6} anharmonicity lifts the instability observed close to $\Gamma$ in the harmonic case.}
    \label{sscha-phonon}
\end{figure}

\begin{figure}[tbh!]
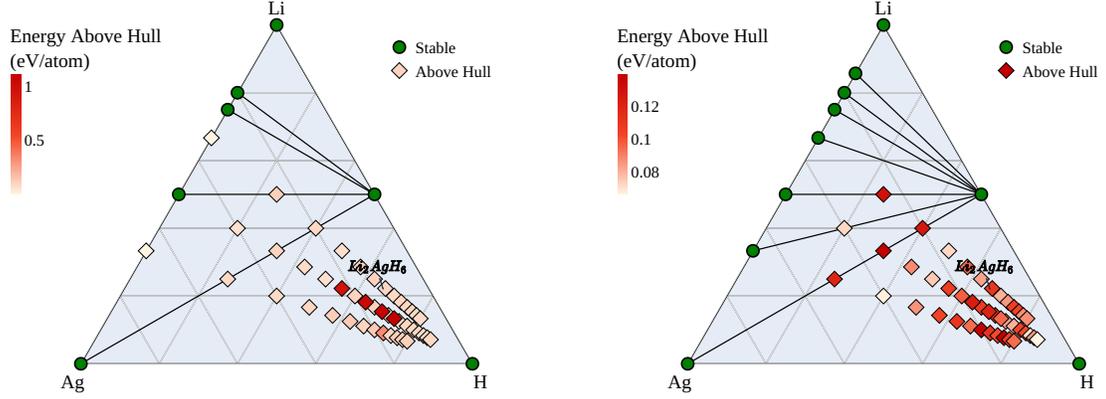

    \centering
    \includegraphics[width=0.43\columnwidth, trim={0.55cm 0 2.2cm 0},clip]{img/phase_diagram_25Pa.pdf}
    \hspace{0.1\columnwidth} \includegraphics[width=0.43\columnwidth, trim={0.55cm 0 2.2cm 0},clip]{img/phase_diagram_50GPa.pdf}\\
    \vspace{-0.3cm}
    \caption{The convex hull of the Li--Ag--H system at 25~GPa (left) and 50~GPa (right). Circles represent thermodynamically stable structures, and diamonds represent metastable structures.}
    \label{}
\end{figure}
\newpage
\bibliographystyle{naturemag}
\bibliography{references.bib}